# A Sacrificial Magnet Concept for Field Dependent Surface Science Studies

*Danyang Liu[1], Jens Oppliger[1], Aleš Cahlík[1], Catherine Witteveen[1,2], Fabian O. von Rohr[2], and Fabian Donat Natterer[1]\**

1. Department of Physics, University of Zurich, Winterthurerstrasse 190, CH-8057 Zurich, Switzerland
2. Department of Quantum Matter Physics, University of Geneva, 24 Quai Ernest-Ansermet, CH-1211 Geneva, Switzerland

*\*fabian.natterer@uzh.ch*

*We demonstrate a straightforward approach to integrate a magnetic field into a low-temperature scanning tunneling microscope (STM) by adhering an NdFeB permanent magnet to a magnetizable sample plate. To render our magnet concept compatible with high-temperature sample cleaning procedures, we make the irreversible demagnetization of the magnet a central part of our preparation cycle. After sacrificing the magnet by heating it above its Curie temperature, we use a transfer tool to attach a new magnet in-situ prior to transferring the sample into the STM. We characterize the magnetic field created by the magnet using the Abrikosov vortex lattice of superconducting $NbSe_2$. Excellent agreement between the distance dependent magnetic fields from experiments and simulations allows us to predict the magnitude and orientation of magnetic flux at any location with respect to the magnet and the sample plate. Our concept is an accessible solution for field-dependent surface science studies that require fields in the range of up to 400 mT and otherwise detrimental heating procedures.*

- *Accessible magnetic field generation*
- *Selectable field strength and orientation*
- *Compatible with high-temperature sample preparation*

**\*Method details**

Concept

The life cycle of our sacrificial magnet concept for surface science studies is described in Figure 1. Our approach consolidates the requirements of high temperature during the sample preparation and the constraints given by the low Curie temperature ($T_C$) of a permanent magnet, as indicated in panel (a). The heating of a sample above $T_C$ is first carried out without a magnet on the sample plate. After the sample heating procedure is completed, we attach a commercial Neodymium-Iron-Boron (NdFeB)(SM) magnet onto a magnetizable sample plate (Steel StW 22) using a magnet transfer tool [see panel (b)] made from the same material as the sample plate. The transfer of the magnet from the tool to the sample plate works because the contact area between magnet and transfer-tool is smaller than between magnet and sample plate (contact area ratio ~1/5). Consequently, the magnet preferably adheres to the sample plate, which is then transferred into the scanning tunneling microscope (STM) for experiments [panel (c)]. Prior to their transfer, we clean the permanent magnets using acetone and ethanol and stick them on the transfer tool. In our system, the transfer tool is connected to a linear transporter that is ordinarily used for sample transfer between a fast entry loadlock and the preparation chamber. The adaptation of the magnet transfer routine therefore requires no change to our ultrahigh-vacuum system and only the fabrication of the transfer tool. After the STM experiments, we conclude the cycle by dropping the magnet [panel (d)] because our approach specifically accepts the irreversible loss of its magnetization during the sample preparation for $T>T_C$.

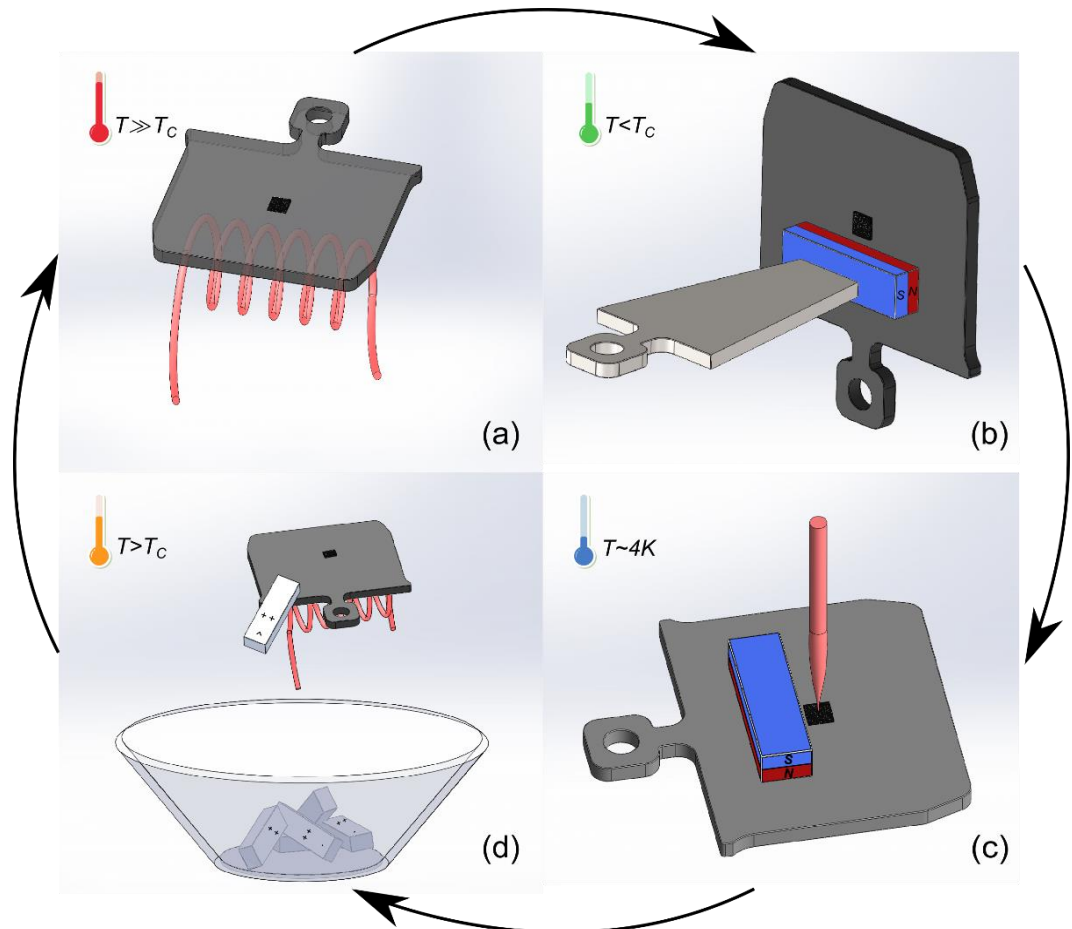

*Figure 1: **The life cycle of a sacrificial magnet.** (a) Annealing the sample is a common step in the sample preparation of surface science experiments, which can include heating of the sample far above the Curie temperature ($T_C$) of a permanent magnet. (b) After the sample-heating is concluded ($T<T_C$), a permanent magnet is attached onto the magnetizable sample plate. (c) The sample is transferred into the STM with the attached magnet providing a magnetic field for the experiment. (d) The magnet is sacrificed by heating the sample plate above $T_C$, leading to its irreversible demagnetization and drop into a collection basket.*

Evaluation

We test our magnet transfer concept using a commercial low-temperature STM system (Createc) and quantify the magnetic field in the vicinity of the magnet to evaluate the spatial flux density for surface studies, as described in the following.

We first get a sense for $T_C$ by heating several NdFeB magnets *ex-situ* until they drop off their steel plates around a temperature of about 400 °C. This shows that the magnet is damaged by a lower temperature than would be used in the preparation of typical metal surfaces, *i.e.* 577 °C for Au(111) [1], emphasizing the need to separate the magnet and the sample plate during sample preparation. Even for lower-annealing temperatures, used to remove water adsorbates or for curing of epoxy adhesives, a permanently connected NdFeB magnet would be at risk of losing part of its magnetization [2].

Next, we test the transfer of the NdFeB magnet from the transfer tool onto a sample plate and its disposal using the regular heating stage on the manipulator, see Figure S1. We can place the magnet at a selected distance with respect to an already mounted single crystal sample. To control the drop-off, we rotate the manipulator with the heater stage overhead to aim it at a stainless-steel basket that we have mounted inside our chamber. Once the temperature exceeds $T_C$, the magnet irreversibly drops into the basket, clearing the sample for an additional heat treatment or sample preparation steps.

To evaluate the magnetic field distribution at the locations where an actual sample could be mounted, we place a NbSe$_2$ single crystal in vicinity of a 10×3×2 mm$^3$ NdFeB magnet (Figure 2). NbSe$_2$ is a type-II superconductor with critical temperature of 7.2 K [3] that exhibits the well-known Abrikosov flux vortex-lattice when subjected to an external magnetic field in the superconducting state [4]. As each vortex is exactly penetrated by one flux quantum $\Phi_0 = \frac{h}{2e} = 2.067 \times 10^{-15}\, Tm^2$, the vortex-lattice parameter can be used to precisely measure the applied magnetic field [4,5]. We therefore determine the mean vortex separation $d_v$ from closed loop conductance scans at a bias voltage providing a good contrast between superconducting [Figure 2(a) purple spectrum] and normal [Figure 2(a) blue spectrum] state to evaluate the magnetic field for a triangular vortex lattice following $B = \frac{\Phi_0}{S} = \frac{2\Phi_0}{\sqrt{3}d_v^2}$ [5]. Having a negligible in-plane component [see Figure 3(a)], $d_v$ characterizes the vertical component $B_\perp$ of the magnetic field generated by the permanent magnet. We investigate $d_v$ for different macroscopic distances *L* to the magnet edge for which we use optical images of the STM location to determine *L* (Figure 2 bottom). We also use deliberately added scratches on the sample holder to help determine the tip position, see Figure S2. Panels (a)-(d) show conductance maps of the triangular vortex lattices measured at 4 tip-locations and the corresponding optical images of the tip-position. As expected, we see a larger vortex separation when the tip is farther away from the magnet, corresponding to a reduction of the flux density from about 350 mT to 150 mT.

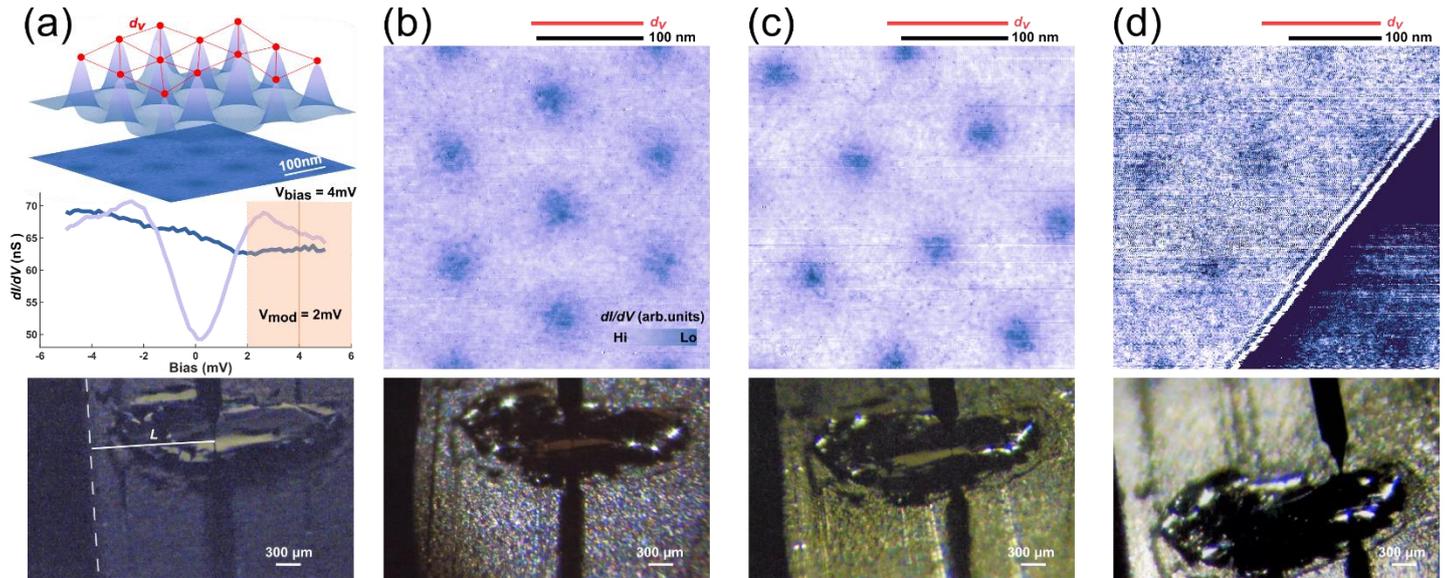

*Figure 2: **Magnetic field measurement using Abrikosov vortex lattices in NbSe$_2$.** (a) The locations of the flux vortices are determined by 2D Gaussian fits to the extrema of the dI/dV maps, allowing us to evaluate the mean vortex separation $d_v$, here measured for L = (1.40 ± 0.08) mm, $d_v$ = (100 ± 3) nm. Representative spectra inside (blue) and outside (purple) the vortex. Vortex lattice measured for **(b)** L = (1.62 ± 0.08) mm, $d_v$ = (105 ± 4) nm, **(c)** L = (1.94 ± 0.08) mm, $d_v$ = (113 ± 5) nm, **(d)** L= (2.45 ± 0.08) mm, $d_v$ = (128 ± 5) nm. The tilted line is a NbSe$_2$ step edge. **(Lower panels)** optical images used to evaluate the distance L between tip and magnet. (T = 4 K, $V_{bias}$ = 4 mV, I = 500 pA, $V_{mod}$ = 2 mV, 887 Hz.)*

In order to predict the magnetic field that would be achieved at the surface of thicker samples, we model the magnet and sample plate geometry for the same dimensions that were characterized by our experiment using the software *Finite Element Method Magnetics* (FEMM) [6]. Figure 3(a) shows the steel sample plate ($\mu_r = \mu_{rFe}$), the NdFeB permanent magnet, the locations *L* (red dots) at which we measure the Abrikosov vortex lattices, and the field-distribution that we obtain from the FEMM simulation. The bands below the simulation curves in Figure 3(b) account for a maximal 14% reduction of the flux density that is associated with a temperature dependent phase transition in NdFeB at about 135 K [4] in which its easy-axis is canted by maximally 30°. The simulation (blue line) in panel (b) was evaluated at a height of 100 µm above the sample-holder which corresponds to the approximate thickness of our NbSe$_2$ crystal. It shows an excellent agreement with our measurements (red dots). With the properly validated calibration, we simulate the field-distribution at 2 mm above the sample plate (pink line) which corresponds to the height of our single crystal substrates, such as Au(111), Ag(100), and Cu$_3$Au(111) [8]. As already visible from the flux-lines in panel (a), the field lines at 2 mm above the sample plate are more tilted in proximity to the magnet. The magnetic flux orientation varies strongly with the distance from the magnet edge, which we further evaluate in panel (c). This demonstrates the possibility to conduct field dependent studies using mostly in-plane to mostly out-of-plane alignment of the magnetic flux on the same surface by just laterally moving the tip to a new location.

When we simulate the magnetic field for a NdFeB magnet mounted on a non-magnetic substrate ($\mu_r = 1$), we find a lower flux density [dashed blue in Figure 3(b)] at the locations of the original experiment, showing the influence of the steel plate in guiding the magnetic field-lines. Shaping a magnetizable sample plate and therefore the field-distribution can be easily carried out to tailor the field for an experiment. Similar to pole-shoes in electromagnets, the field-lines can be bundled, and their density maximized at a desired location. With slight changes to the sample-plate geometry, it is possible to achieve flux densities of more than 500 mT, making the sample plate a highly versatile design parameter.

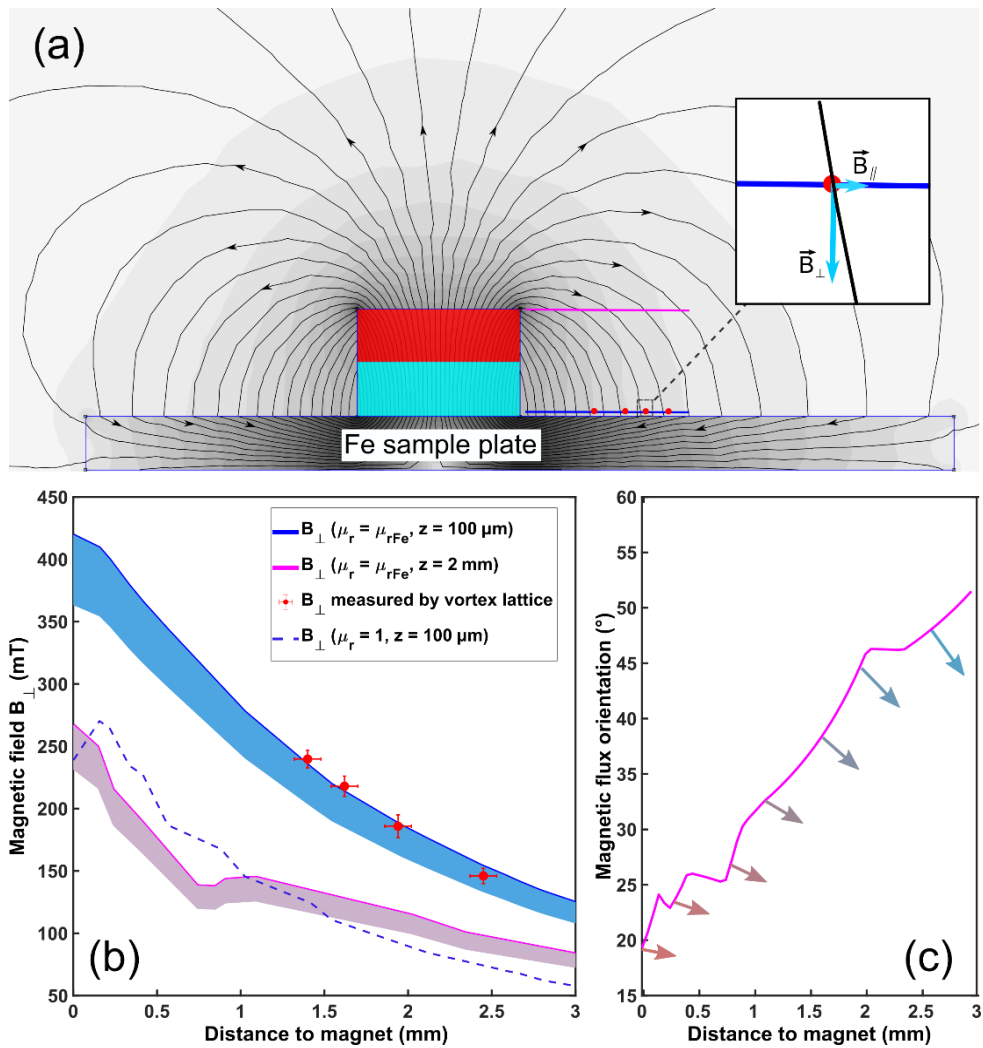

*Figure 3: **Spatial magnetic field gradient**. (a) 2D simulation of magnetic flux density using the software FEMM. (b) Magnetic field as a function of the distance to the magnet, showing good agreement between the measured (red dots) and simulated field (solid blue line) for a steel plate sample holder ($\mu_r = \mu_{rFe}$). Based on this agreement, we can predict the field at different heights with respect to the sample plate (pink solid line evaluated at 2 mm). Using a non-magnetic ($\mu_r = 1$) sample holder yields a much lower field as shown in the dashed blue field dependence. (c) The angle of the magnetic flux evaluated for the pink line in (b), showing gradually decreasing tilt.*

In conclusion, we have demonstrated and characterized a versatile sacrificial magnet concept that can be implemented in a straightforward fashion with existing vacuum systems. We find that there is an excellent agreement between the measured field-values and finite element modeling. This observation confirms that the presented advanced design can be tailored to the desired field-distribution, angle, and amplitude. Our method is compatible with high sample temperatures that would irreversibly demagnetize NdFeB magnets. We anticipate the introduction of creative designs for the layout of the sample-plates used for advanced surface science studies.

**Acknowledgements:**
F.D.N. thanks the Swiss National Science Foundation (PP00P2_176866 and 200021_200639) and ONR (N00014-20-1-2352) for generous support. D.L. thanks UZH Forschungskredit (FK-20-093). C.W. and F.O.vR acknowledge the support from the Swiss National Science Foundation (PCEFP2_194183) and by the Swedish Research Council (VR) through a neutron project grant Dnr. 2016-06955.

*Author Contributions:*
F.D.N. and D.L. conceived the project. D.L. and F.D.N. wrote the manuscript. D.L., A.C., and J.O. measured the data. D.L., J.O., and F.D.N. analyzed the data. D.L. and J.O. prepared the samples. C.W. and F.O.vR. synthesized the NbSe$_2$ samples. F.D.N. supervised the project.

**Declaration of interests:**
The authors declare that they have no known competing financial interests or personal relationships that could have appeared to influence the work reported in this paper.

**Supplementary material *and* Additional information:**

*Sample preparation*

We purchased commercial NdFeB magnets (supermagnete.ch: Q-10-03-02-HN) and wiped them with acetone and ethanol. The NbSe$_2$ flake is taken from the same batch of crystals described in Ref. [3] that were grown by the chemical vapor transport method (ibid). We glue a single crystal of NbSe$_2$ together with a cleaving post onto the sample holder, attach a magnet next to it, and transfer it without any further heat-treatment into our ultra-high vacuum system. Prior to transferring the sample into the 4 K cooled STM, we cleave the NbSe$_2$ crystal in-vacuum. For STM measurements, we use a tip made from electrochemically etched W that we treated by gently plunging it into an Au(111) sample until the apex is atomically sharp, which we verify by scanning across Au step edges. We record point and closed-loop spectroscopy using a conventional lock-in technique.

*Video Stills of the magnet insertion and sacrifice*

In Figure S1 we demonstrate the attachment of a NdFeB permanent magnet onto the sample plate next to the Cu$_3$Au(111) crystal using the transfer tool, as shown in panel (a) and (b). The sacrifice of the magnet into the basked can be seen in panels (c) and (d), showing the heating of the sample and the magnet's drop off into the collection basket.

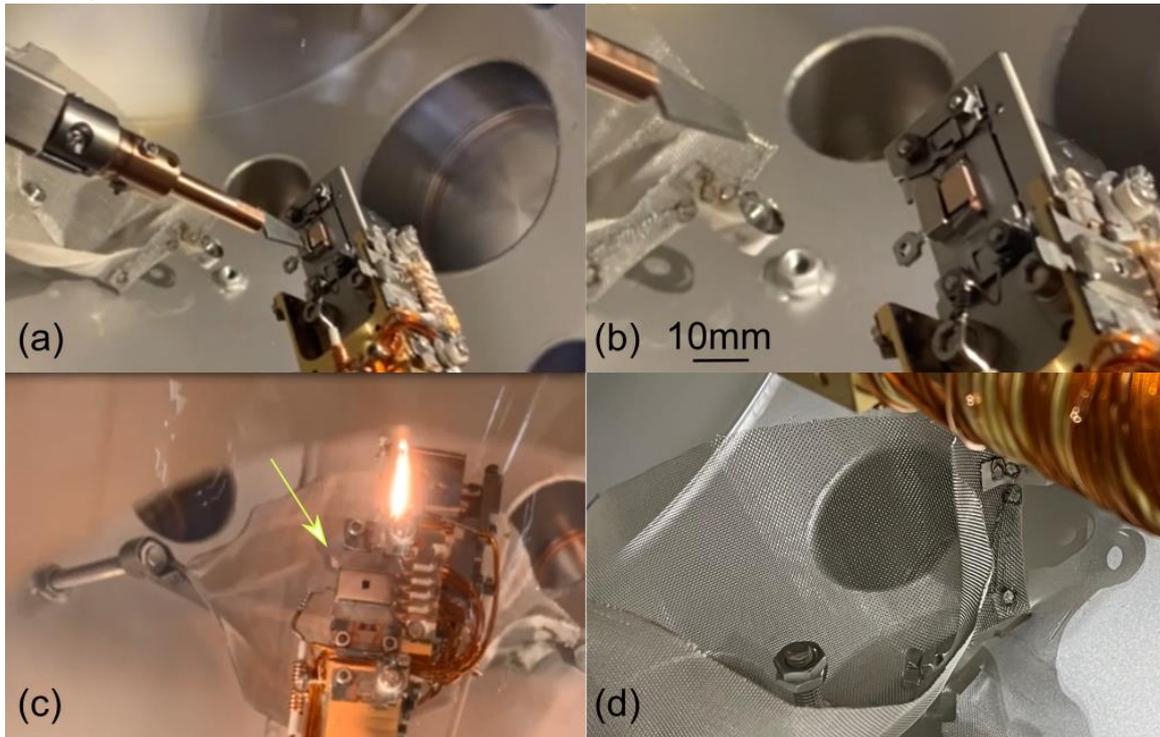

*Figure S1: **(a)** Permanent magnet attachment using the transfer tool. **(b)** Retraction of the transfer tool and attached magnet next to a Cu3Au crystal. **(c)** The magnet drops by heating it above T$_C$. **(d)** Collection basket for the sacrificed magnets.*

*Determine the distance L from the tip to the magnet edge*

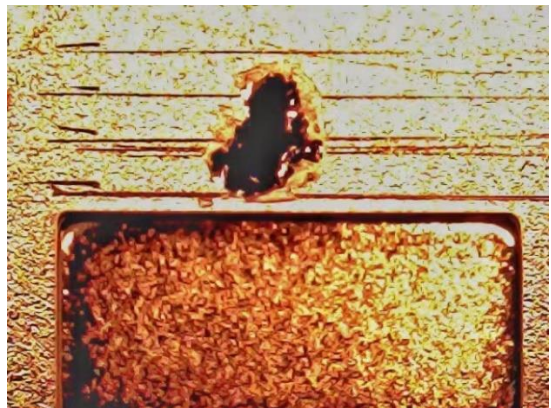

*Figure S2: Scratches on the sample holder to help determine the tip position L with respect to the magnet edge. The long side of the magnet sets the scale to 10 mm.*

**\*References:**